\documentclass[]{spie}  

\usepackage{amsmath,amsfonts,amssymb}
\usepackage{graphicx}
\usepackage[colorlinks=true, allcolors=blue]{hyperref}

	\usepackage{amsmath}
	\usepackage{mathrsfs} 
	\usepackage{amssymb}
	\usepackage{amsfonts}
	\usepackage{graphicx}  
	\usepackage{setspace}
	\usepackage{cancel}
    \usepackage[left=0.875in,right=0.875in,top=1in,bottom=1.25in]{geometry}
	\usepackage{enumitem}
	\usepackage{etoolbox}
	\usepackage{siunitx}	
	\usepackage{multirow,bigdelim}
	\usepackage{textcomp}
	\usepackage{float}
    \usepackage{upgreek}
	\usepackage{color, colortbl} 
	\usepackage{booktabs} 
	\usepackage{caption}
	\usepackage[font=singlespacing]{caption}
	\usepackage{wasysym}
	\usepackage{pdflscape}
	\usepackage{afterpage}
	\usepackage{url}
    \usepackage[perpage]{footmisc} 
	\usepackage{lscape} 
	\usepackage{subcaption} 
	\usepackage{mathtools}          

	\AtBeginEnvironment{quote}{\singlespacing\small\it}
	
	 \newcommand{\arcsec}{$^{\prime\prime \xspace}$}	
    \newcommand{\arcmin}{$^{\prime \xspace}$}
	\usepackage{xspace}
    \newcommand{\um}{$\upmu$m\xspace}	
	
	\renewcommand{\deg}{$^{\circ}$ }

	\newcommand{\planck}{\emph{Planck }}
	\setlist[enumerate]{label*=\arabic*.}
	\setlist[enumerate,1]{label*=\arabic*.,font=\bfseries,before=\bfseries}
	\setlist[enumerate,2]{label*=\arabic*.,font=\normalfont,before=							\normalfont}
	\setlistdepth{9}

\title{Preflight characterization of the BLAST-TNG receiver and detector arrays}

\author[a]{Nathan P. Lourie}
\author[i]{Peter A. R. Ade}
\author[a]{Francisco E. Angile}
\author[g]{Peter C. Ashton}
\author[b]{Jason E. Austermann}
\author[a]{Mark J. Devlin}
\author[b]{Bradley Dober}
\author[f]{Nicholas Galitzki}
\author[b]{Jiansong Gao}
\author[d]{Sam Gordon}
\author[d]{Christopher E. Groppi}
\author[a]{Jeffrey Klein}
\author[b]{Gene C. Hilton}
\author[b]{Johannes Hubmayr}
\author[c]{Dale Li}
\author[a]{Ian Lowe }
\author[d]{Hamdi Mani}
\author[d]{Philip Mauskopf}
\author[b]{Christopher M. McKenney}
\author[a]{Federico Nati}
\author[h]{Giles Novak}
\author[i]{Enzo Pascale}
\author[i]{Giampaolo Pisano}
\author[d]{Adrian Sinclair}
\author[e]{Juan D. Soler}
\author[i]{Carole Tucker}
\author[b]{Joel Ullom}
\author[b]{Michael Vissers}
\author[h]{Paul A Williams}
\affil[a]{University of Pennsylvania, 209 South 33rd St, Philadelphia, USA}
\affil[b]{National Institute of Standards and Technology, 325 Broadway, Boulder, USA}
\affil[c]{SLAC National Accelerator Lab, 2575 Sand Hill Rd, USA}
\affil[d]{School of Earth and Space Exploration, Arizona State University, Tempe, USA}
\affil[e]{Max-Planck-Institute for Astronomy, Konigstuhl 17, Heidelberg, Germany.}
\affil[f]{University of California San Diego, 9500 Gilman Dr, La Jolla, USA}
\affil[g]{University of California - Berkeley, Lawrence Berkeley National Laboratory, 1 Cyclotron Rd, CA, USA}
\affil[h]{Center  for  Interdisciplinary  Exploration  and  Research  in Astrophysics  and  Department  of  Physics  \&  Astronomy,  Northwestern University,  2145 Sheridan  Road, USA}
\affil[i]{Cardiff University, Cardiff CF10 3AT, UK}

\authorinfo{Further author information: (Send correspondence to N.P.L.)\\N.P.L.: E-mail: nlourie@sas.upenn.edu, Telephone: 1 215 573 5445}

\begin{document} 
\maketitle

\begin{abstract}
The Next Generation Balloon-borne Large Aperture Submillimeter Telescope (BLAST-TNG) is a submillimeter mapping experiment planned for a 28 day long-duration balloon (LDB) flight from McMurdo Station, Antarctica during the 2018-2019 season.  BLAST-TNG will detect submillimeter polarized interstellar dust emission, tracing magnetic fields in galactic molecular clouds. BLAST-TNG will be the first polarimeter with the sensitivity and resolution to probe the $\sim$0.1 parsec-scale features that are critical to understanding the origin of structures in the interstellar medium. 

BLAST-TNG features three detector arrays operating at wavelengths of 250, 350, and 500 \um (1200, 857, and 600 GHz) comprised of 918, 469, and 272 dual-polarization pixels, respectively. Each pixel is made up of two crossed microwave kinetic inductance detectors (MKIDs). These arrays are cooled to 275 mK in a cryogenic receiver. Each MKID has a different resonant frequency, allowing hundreds of resonators to be read out on a single transmission line. This inherent ability to be frequency-domain multiplexed simplifies the cryogenic readout hardware, but requires careful optical testing to map out the physical location of each resonator on the focal plane. Receiver-level optical testing was carried out using both a cryogenic source mounted to a movable xy-stage with a shutter, and a beam-filling, heated blackbody source able to provide a 10-50 $^\circ$C temperature chop. The focal plane array noise properties, responsivity, polarization efficiency, instrumental polarization were measured. We present the preflight characterization of the BLAST-TNG cryogenic system and array-level optical testing of the MKID detector arrays in the flight receiver.

\end{abstract}


\section{INTRODUCTION}
\label{section:intro}  

The Next Generation Balloon-borne Large Aperture Submillimeter Telescope (BLAST-TNG) is a submillimeter mapping experiment which features three microwave kinetic inductance detector (MKID) arrays operating over 30\% bandwidths centered at 250, 350, and 500 \um (1200, 857, and 600 GHz). These highly-multiplexed, high-sensitivity arrays, featuring 918, 469, and 272 dual-polarization pixels, for a total of 3,318 detectors, are coupled to a 2.5 m diameter primary mirror and a cryogenic optical system providing diffraction-limited resolution of 30\arcsec, 41\arcsec, and 50\arcsec \space respectively. The arrays are cooled to $\sim$275 mK in a liquid-helium-cooled cryogenic receiver which will enable observations over the course of a 28-day stratospheric balloon flight from McMurdo Station in Antarctica as part of NASA's long-duration-balloon (LDB) program, planned for the 2018/2019 winter campaign. BLAST-TNG is the successor to the BLASTPol and BLAST balloon-borne experiments which flew five times between 2005 and 2012\cite{enzo_blast,fissel_blastpol}.

Achieving diffraction-limited, sub-arcminute resolution and telescope pointing accuracy is one of the highest priorities for the success of the BLAST-TNG mission. Although the science goals of BLAST-TNG are similar to the 2012 BLASTPol mission, most of the major instrument systems have been rebuilt and improved since the last flight. A new 2.5 m aperture Cassegrain telescope, featuring a lightweight composite carbon fiber reinforced polymer (CFRP) primary mirror designed and built by Alliance Spacesystems,\footnote{4398 Corporate Center Dr, Los Alamitos, CA 90720} will enable an increase in resolution to 30\arcsec \space at 250 \um,  from BLASTPol's 2.5\arcmin \space at the same band. With improved detector sensitivity and a increase in detector count by a factor of 12, we expect BLAST-TNG will have more than six times the mapping speed of BLASTPol. The new cryostat has demonstrated a 28 day hold-time, enabling observations of many more targets at greater depth than were possible during the $\sim$13 day BLASTPol flight in 2012.

The primary science goal of BLAST-TNG is to map the polarized thermal emission from galactic interstellar dust around star-forming regions and in the diffuse interstellar medium (ISM). These maps will yield $\sim$250,000 polarization vectors on the sky, allowing us to explore correlations between the magnetic field dispersion, polarization fraction, cloud temperature, and column density. Quantifying the relationships between these variables over a large sample of clouds will yield testable relationships which can be fed back into numerical simulations. The \planck satellite has observed strong correlations between the orientation of Galactic magnetic fields and large-scale ISM structures \cite{planck_XXXII}, as well as the interior of giant molecular clouds (GMCs)\cite{planck_XXX}. While BLASTPol was able to observe the magnetic fields within GMCs at higher resolution than \planck \cite{fissel_blastpol,soler_blastpol}, BLAST-TNG will be the first experiment to probe the fields within the characteristic filamentary structures within GMCs observed by \emph{Herschel} \cite{hill}. Combining the BLAST-TNG data with molecular cloud simulations, \cite{soler_sims_2013}  and numerical models of dust emission \cite{guillet_2018} and  grain properties, \cite{draine_hensley} will give unprecedented insight into the interplay between the gravitational, turbulent and magnetic field contributions to star and cloud formation, as well as the physics of grain alignment and mass flow within the interstellar medium.

Polarized dust emission is also the dominant foreground for observations of the cosmic microwave background (CMB). Characterization of these foregrounds is one of the most important requirements in the search for the gravitational wave signature of cosmic inflation \cite{cmb_s4}. While the power spectrum from polarized dust foregrounds is thought to be lowest at small angular scales, there is limited high-resolution observational data of the diffuse ISM \cite{planck_XXX,caldwell_EEBB}.  BLAST-TNG will be able to make the deepest maps to date of the dust emission in the types of dark, diffuse regions of the sky favored by state of the art CMB polarization experiments. BLAST-TNG will probe angular scales not well-characterized to date, and explore correlations between diffuse dust emission and structures in the cold neutral medium \cite{ghosh} at submillimeter wavelengths where the intensity of the thermal dust signal dominates.  With its high pixel count and photon-noise-limited detectors, BLAST-TNG will produce maps of diffuse ISM with higher fidelity than the highest frequency \emph{Planck} polarization maps at 353 GHz.

\section{Cryogenic Receiver}
\label{section:receiver_overview}

Success of BLAST-TNG's observational goals relies on making sensitive and stable observations of a large sample of molecular clouds, and a varied sample of regions of the ISM with different densities and radiative environments. This is enabled primarily through by the high-sensitivity MKID detector arrays \cite{chris_kids_spie,vissers_kids_spie}, careful control of polarization systematics, and crucially, a cryogenic receiver that which operates autonomously for the full 28 day flight.

\subsection{Cold Optics}
\label{section:cold_optics}
\begin{figure}[ht]
\includegraphics[width=\linewidth]{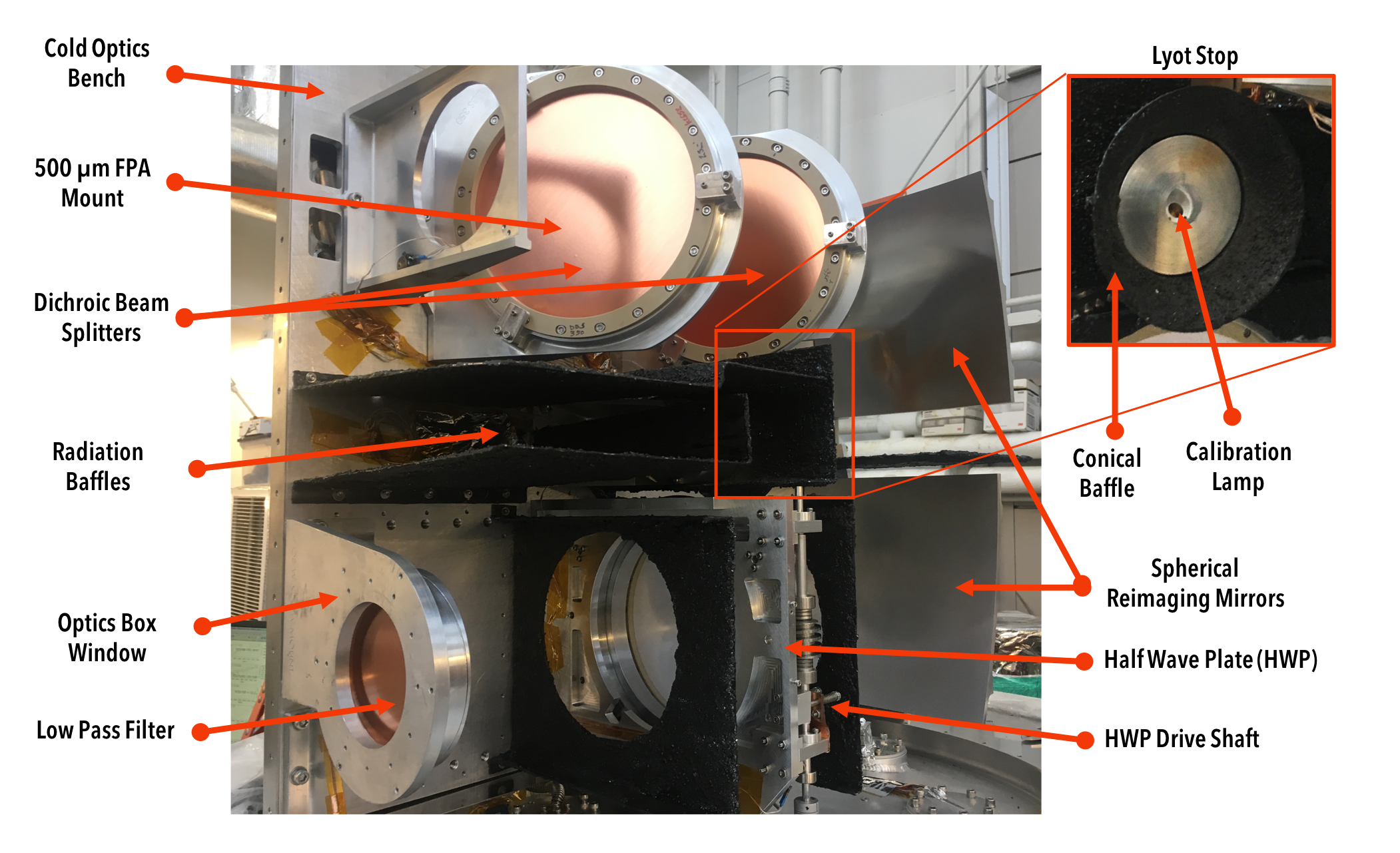}
\caption[Photograph of the Optics Bench] {Photograph of the BLAST-TNG 4 K reimaging optics  with critical components labeled.}
\label{figure:optics_bench}
\end{figure}

The cryogenic receiver encloses a series of cold re-imaging optics arranged in a modified Offner relay configuration. A similar configuration was flown in the BLAST/BLASTPol optics box. The design features a cold Lyot stop at the image of the primary mirror which significantly reduces the optical loading on the detectors from stray light.  The optics bench, shown in Fig. \ref{figure:optics_bench}, cools the 4 K reimaging optics, and supports the band-defining filters which split the telescope beam to the three focal plane arrays. Details of the optical design can be found in Lourie, et. al\cite{lourie_telescope}, and in refs. \citenum{tyr_blasttng_spie,brad_blast_spie}.

\subsection{Cryostat Design}
\label{section:cryostat}  

To simplify the mechanical design and minimize the number of pressure vessels, the cryostat is based on liquid helium-only system. A 250-L liquid helium tank cools the optics and cold electronics to 4 K, and backs the operation of the sub-Kelvin refrigeration system, described in further detail in Section \ref{section:fridges}, and in Galitzki, et. al. \cite{tyr_blasttng_spie}.

\begin{figure}[ht]
\centering
\includegraphics[width=.8\linewidth]{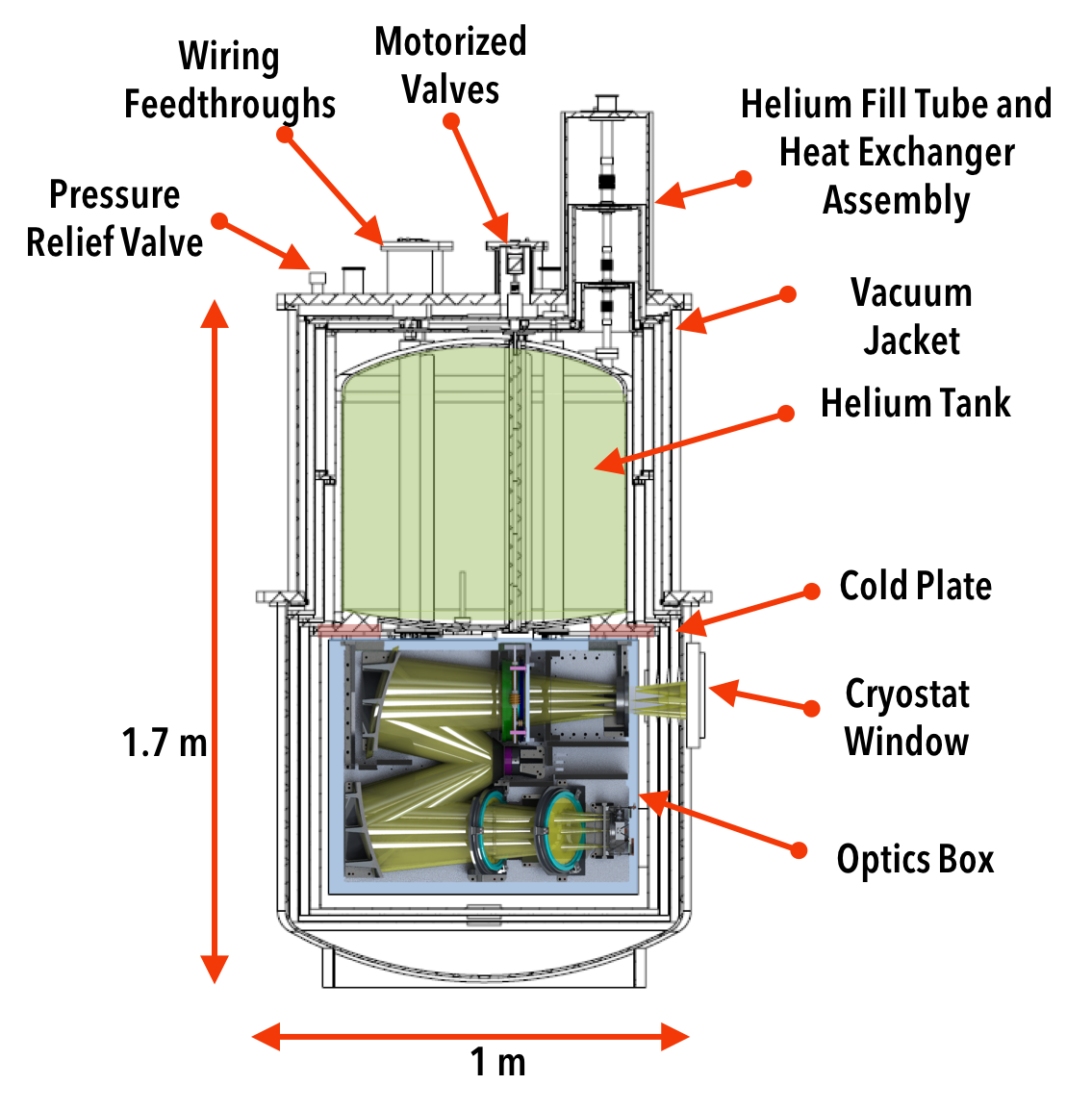}
\caption[Cross-section Render of Cryostat] {Cross-section render of the BLAST-TNG cryostat with critical components labeled.}
\label{figure:cryo_cutaway}
\end{figure}

The 4 K cold plate is integrated into the tank, and forms the lower cap of the liquid helium dewar as shown in Fig. \ref{figure:cryo_cutaway}. The cold plate has a domed center to maximize structural rigidity while reducing mass. The optics bench is bolted to the thick rim of the tank and located with precision alignment features machined around the perimeter. By mounting the optics to the perimeter, the optics are isolated from pressure-induced bowing at the center of plate. The 4 K optics cavity is enclosed by an 1100-series aluminum shroud. The interior of this shroud is coated with an absorptive coating made from Stycast 2850-MT/Cat 23LV\footnote{Henkel Adhesives, North America} mixed with 10\% by mass of powdered charcoal\footnote{General Pencil Company, Inc. Redwood City, CA} to absorb stray in-band light as well infrared radiation to prevent indirect loading of the sub-Kelvin components\cite{persky_black_surfaces}.

All housekeeping electronics are accessed via the top lid of the cryostat. A series of six pass-through pipes are welded into the liquid tank to allow wiring, coaxial connections from the detector focal plane readout, and axles from the cryogenic actuators to be passed directly from the cold-plate to the top of the cryostat. These pass-throughs can be accessed by  removing the top lids of the vacuum shell and vapor-cooled shields, so that making changes to the wiring harnesses does not require disassembly or removal of all of the cryostat shells. 

Two intermediate thermal shrouds made of 1100-series aluminum enclose the 4 K volume to reduce the conductive and radiative loading on the liquid helium bath. As the liquid helium boils, the vapor is forced through two copper spiral heat exchangers \cite{tyr_thesis}, one bolted to each of these intermediate vapor-cooled stages (VCS), cooling the first intermediate stage (VCS1) to 40 K and the second stage (VCS2) to 140 K. The entrance and exit apertures of the spiral heat exchangers sit within the helium fill port, which is the only port attached to the helium tank. Two spring-loaded PTFE plugs seal the fill port at each VCS stage, allowing vapor pressure in the tank to build to 25 mbar above atmospheric pressure and force the cold helium gas from the cryogen boil-off through the heat exchangers\cite{tyr_thesis}. These plugs are removed during cryogen transfers. A TAVCO\footnote{TAVCO Sales \& Service Company, Inc. Gilbert, AZ} 1-atm absolute pressure valve regulates the pressure at the outlet of the VCS2 heat exchanger. The 4, 40, 140, and 300 K stages are separated by G10 fiberglass cylinders with wall thicknesses of 0.5, 1.0, and 1.6 mm respectively. The G10 cylinders are assembled from epoxy-coated woven fiberglass, and are assembled such that the warp of the G10 fibers is oriented circumferentially around the cylinders which reduces the effective thermal conductivity of the supports \cite{g10}. The VCS stages are a highly coupled system, and the cooling power of the heat exchangers is proportional to the boiloff rate of the helium bath, providing negative thermal feedback to the 4 K stage. The VCS typically reach \textpm 5 K of their equilibrium temperatures within 48 hours of the initial liquid helium transfer and reach equilibrium temperatures stable to \textless 1 K within 4 days.

If not properly controlled, infrared loading through the cryostat window can dominate the loading at each stage. A series of metal-mesh band-defining and thermal/infrared blocking filters, and low-pass band-defining filters reflects infrared light back out of the cryostat, while passing submillimeter wavelengths through to the cold optics and FPAs. The filter arrangement at each temperature stage has been adjusted between ``light runs" with the window installed, in order to optimize the infrared rejection and maximize in-band transmission. The radiative load on the 4 K stage is particularly sensitive to the filter arrangement at the two VCS. The low-pass filters reject out-of-band submillimeter light, but absorb infrared wavelengths, which must be rejected  earlier in the filter stack. Where allowed by space constraints, the filters are tipped at opposing angles to reduce Fabry-Perot resonances and multiple reflections.

\subsection{Sub-Kelvin Refrigeration System}
\label{section:fridges}  
The BLAST-TNG focal plane arrays are cooled to $\sim$275 mK via a closed-cycle $^3$He sorption refrigerator, backed by a $\sim$1 K superfluid, pumped $^4$He volume (the ``pumped pot"), which draws liquid helium from the main liquid helium tank. The $^3$He refrigerator is a copy of that flown on the BLASTPol and BLAST experiments\cite{enzo_blast}, and built for the MUSTANG instrument on the Green Bank Telescope\cite{he3_fridge}.

\begin{figure}[ht]
\begin{center}
\includegraphics[width=\textwidth]{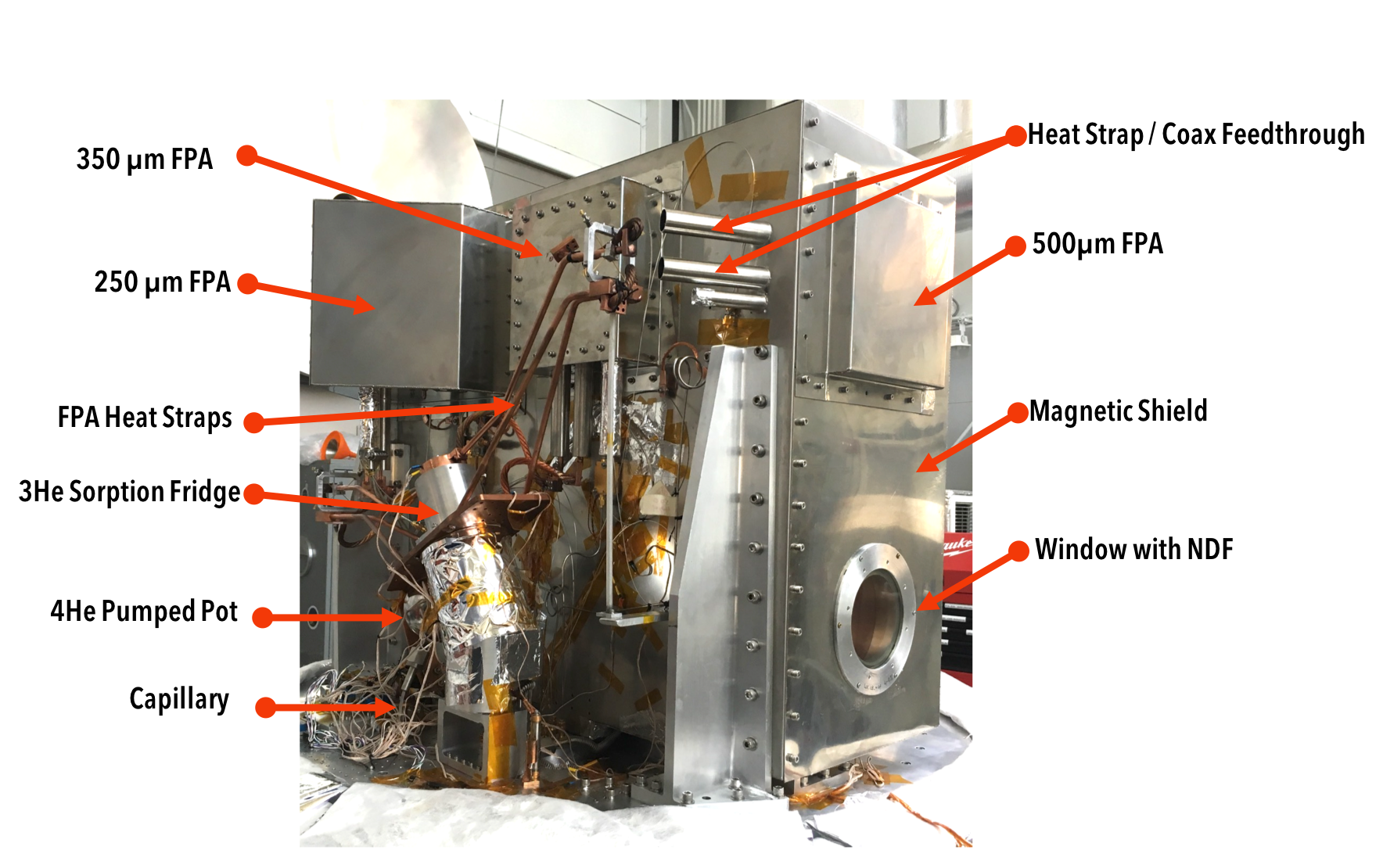}
\caption[Cold Optics and Refrigeration System] {Photograph of the BLAST-TNG cold plate, showing the cold optics box covered by the magnetic shielding, and the sub-Kelvin refrigeration components.}
\end{center}
\label{figure:coldplate_redline}
\end{figure}%

The geometry of the BLAST-TNG superfluid system is designed to minimize the consumption of liquid helium during operation, and run as cold as possible to reduce loading on the FPAs. The flow of liquid helium into the pumped pot is regulated by a 0.25 mm diameter rate-limiting capillary, and can be turned on and off by a cryogenic valve\footnote{Swagelok Solon, OH}. This is contrasted by similar systems in the BLASTPol and SPIDER cryostats in which the pumped-superfluid pot is continually filled by a smaller capillary \cite{enzo_blast,spider_cryo}. While introducing a cryogenic valve increases the complexity and risk of in-flight valve failure, it greatly reduces the consumption of liquid helium over the course of the flight and the larger capillary size reduces the risk clogs or ice-plugs. Additionally, stopping the flow of liquid helium into the pot lowers the base temperature of the system, since the pot will operate at a lower vapor pressure for a given pumping speed. During flight, the pot is pumped to the ambient pressure at $\sim$35 km altitude through a 19 mm diameter pump tube. We expect in-flight operation below 1.3 K, an improvement from the 1.8 K BLASTPol system \cite{elio}. The 4 K valve is actuated via a G10 fiberglass shaft, through a ferrofluidic feed-through\footnote{FerroTec Corporation, Santa Clara, CA}, driven by a geared stepper motor mounted outside the vacuum shell on the top of the cryostat. 

The diameter of the capillary was chosen such that the 200 mL pumped-pot can be completely filled in less than half an hour. While filling, some amount of helium entering the pot will be pumped directly through the pot and into the pump tube rather than collecting. By measuring the flow rate of helium gas through the pump with the 4 K valve open and closed, we find the flow rate of liquid helium through the capillary to be 13.3 mL/min, which collects in the pot at a rate of 8.3 mL/min, a filling efficiency of $\sim$60\%.  With the valve open, the pot temperature rises above 2 K. The capillary diameter is such that if the pot valve were to be stuck in its open position due to a mechanical failure, the flight maximum flight time would be reduced to 13 days before the full helium tank is depleted. 

The helium consumption of the pumped-pot during operation can be quantified by calculating an average equivalent thermal load which would consume the same volume of liquid helium over the course of the flight. The pot is sized such that the $^3$He refrigerator can be recycled without refilling, and in practice must be refilled every $\sim$3 days. By turning off the flow of helium into the pumped-pot when it is full, we reduce the equivalent thermal load compared to BLASTPol by nearly 85\%, from 23 mW to 3.5 mW, even while increasing the capillary diameter from 0.038 to 0.25 mm.

The cryostat and the sub-Kelvin system must be able to operate entirely autonomously. While commanding and communications from the ground will be available during the flight, the telemetry bandwidth is limited and equipment failures could cause contact to be lost completely. Housekeeping thermometry is continuously read out via a combination of custom thermometry bias/demodulation electronics and commercial off-the shelf data acquisition hardware\footnote{LabJack Corporation, Lakewood, CO}. Any time the array temperatures exceed threshold values, the flight computer triggers the pot valve to open and recycle the $^3$He sorption refrigerator. The thermal loading is low enough that lab testing indicates that the sorption refrigerator will have to be cycled only once every every 4-5 days.

\section{Receiver Performance}
\label{section:cryo_performance}
The cryostat, designed at the University of Pennsylvania\cite{tyr_jai} and built by Precision Cryogenic Systems\footnote[1]{Precision Cryogenic Systems, Indianapolis, IN}, was first delivered in October, 2015. Preliminary testing indicated the presence of excess loading on each of the thermal stages. During dark tests with the windows covered at each thermal stage, VCS1/2 ran at 65 K and 165 K respectively, and the loading on the liquid helium bath was $\sim$40\% larger than modeled, corresponding to a shortened 22.5 day hold time. The excess loading was attributed to un-modeled radiative loads from light leaks between thermal stages and inadequate multilayer insulation (MLI) around feedthroughs and fixtures \cite{tyr_blasttng_spie}.

In June, 2017 the cryostat experienced a catastrophic cryogenic failure during a pre-cooling procedure with liquid nitrogen, when an ice plug on the fill port caused an over-pressurization of the liquid cryogen tank rupturing the tank welds. The rupture caused liquid nitrogen to spill into the cavity between the tank and VCS1 where it flash-boiled. The ensuing pressure wave destroyed most of the VCS and 4 K shrouds, the magnetic shielding around the optics box, the focal plane mounts, the plumbing for the sub-Kelvin refrigeration system and the housekeeping wiring. Crucially, however, the cold optics, the cold plate, $^3$He refrigerator, the heat exchangers, and the vacuum vessel were determined to be undamaged. The cryostat was rebuilt and assembled, and underwent its first dark test in December, 2017. 

The rebuilt BLAST-TNG cryogenic receiver performance has been validated during extensive laboratory testing and has benefited from key redesigns from its initial conception. Rather than hand-cutting and wrapping single layers of polyester-fiber-backed aluminized mylar to form MLI blankets, custom-designed laser-cut 10-layer blankets of Coolcat 2 NW\footnote[1]{RUAG Space GmbH, Vienna, Austria} were purchased. Layering these blankets provided 10, 20, and 30 layers of aluminized mylar at the around the 4 K, 40 K, and 140 K stages. The laser-cut slits for the various housekeeping components, along with the addition of metallic baffles around the motor axle shafts significantly reduced the light leaks between stages. Dark testing indicates that the excess  loading at each stage has been reduced, and the performance matches the modeled 28 day hold time.

\section{MKID Detector Arrays}
\label{section:arrays}  

The BLAST-TNG detectors are based on arrays of Microwave Kinetic Inductance Detectors (MKIDS). MKIDs are superconducting, lumped element inductive/capacitive (LC) circuits with a resonant frequency and quality factor that is sensitive to changes in incident radiation. Absorbed radiation with enough energy to break Cooper pairs in the circuit causes a change in the kinetic inductance of the device, changing its impedance and causing a shift in the resonant frequency\cite{kids_day}. MKIDs can be highly multiplexed, and arrays can be formed by capacitively coupling multiple resonators to a single transmission line. BLAST-TNG achieves multiplexing factors up to $\sim$1000 (see Table \ref{table:kids}) using a `tile-and-trim' approach in which arrays of resonators with identical inductive absorbers and capacitive elements of multilayer TiN/Ti/TiN films are laid out on a silicon substrate, before trimming the capacitors with a deep-reactive-ion-etch to uniquely tune the resonant frequency of each resonator \cite{chris_kids}. Fabrication errors and wafer non-uniformity can cause displacement of the resonances from their designed frequencies, leading to an ambiguity in the mapping between resonant frequency and physical location on the array. Collisions or overlap between resonances can lead to unusable detectors and reduced yield. The physical location of each resonator was mapped at NIST-Boulder using an custom array of optical light-emiting diodes (LEDs) designed for each FPA \cite{liu_led_mapper,liu_kids}.

\begin{figure}[ht]
\includegraphics[width=\linewidth]{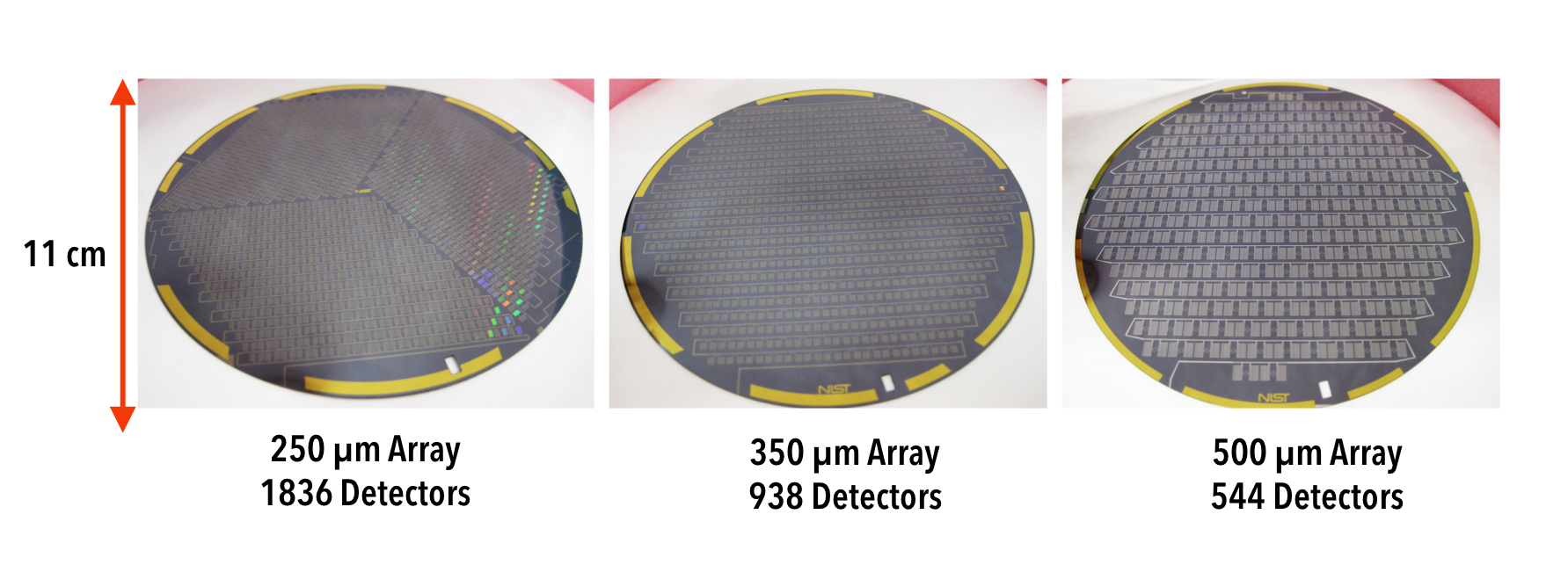}
\caption[FPA Wafers] {Photographs of each of the three focal plane arrays before mounting in their carriers. The photolithographed MKIDs can be seen, along with the meandering transmission line on which they are read out.}
\label{figure:wafers}
\end{figure}

The three BLAST-TNG detector arrays are shown in Fig. \ref{figure:wafers}. The 350 and 500 \um arrays are both read out on a single transmission line, while the 250 \um array is split into three identical rhombus-shaped subarrays, each with its own transmission line. By using the long, thin inductive element of the MKID as the absorber itself, each resonator is sensitive to single linear polarization. Dual-polarization pixels are formed by coupling two single-polarization-sensitive detectors to a single feedhorn, in a crossoverless configuration which achieves less than 3\% cross-polar coupling \cite{brad_kids}. Single-pixel testing with a temperature-controlled blackbody load indicate that the detectors are photon-noise-limited for at their 275 mK operating temperature and 15 pW expected optical loading \cite{hubmayr_kid_photon_noise,chris_kids}. 

\begin{table}[htbp]
  \centering
  \caption{BLAST-TNG Focal Plane Arrays}
    \begin{tabular}{cccc}
    \toprule
    Array Band Center & \multicolumn{1}{l}{Number of Feedhorns} & \multicolumn{1}{l}{Number of MKIDs} & \multicolumn{1}{l}{Multiplexing Factor} \\
    \midrule
    250 \um & 918   & 1836  & 612 \\
    350 \um & 469   & 938   & 938 \\
    500 \um & 272   & 544   & 544 \\
    \bottomrule
    \end{tabular}%
  \label{table:kids}%
\end{table}%

The BLAST-TNG detector arrays are read out using a highly multiplexed, field-programmable gate array (FPGA)-based digital spectrometer. This readout is the first of its kind to have been developed for the second generation Reconfigurable Open Architecture Computing Hardware (ROACH-2) board developed by the CASPER collaboration\cite{casper}. Each readout module includes a ROACH-2 board, MUSIC \cite{music} digital-to-analog and analog-to-digital converter boards (DAC/ADC) and a set of analog radio-frequency (RF) components, which are housed in a custom enclosure. Using firmware designed for BLAST-TNG, a single board is capable of simultaneous readout of over 1000 detectors at a rate of 488 Hz, over 512 MHz of RF bandwidth. Each ROACH-2 module generates a baseband carrier waveform containing the resonant frequencies of each detector. The carrier signal, which is multiplexed on a single coaxial cable, is upconverted to RF and passed to the detector array, where its phase is modulated by the sky signal. The carriers are then amplified by a $\sim$4 K SiGe cryogenic low-noise amplifier (developed at Arizona State University), converted to baseband, and looped back into the ROACH-2, where they are digitized, demodulated, and stored to disk. Five readout modules (three for the 250 \um array, one each for the 350 and 500\um arrays) are mounted in an enclosure mounted on the balloon gondola frame. Details of the readout and pre-flight demonstration are presented in Gordon et al. 2017 \cite{sam_roaches}.

\subsection{Detector Integration}

The first detector tests in the rebuilt BLAST-TNG receiver began in February, 2018. The 350 \um array was installed first and run completely in the dark to characterize the non-optical thermal loading on the array. Each array is mounted on a rigid carbon fiber mount which mounts to the 4 K optics box, and supports the array off of a 1.4 K intercept stage. These mounts conduct less than 1 $\upmu$W per array of thermal power to $^3$He refrigerator\cite{brad_blast_spie}. During dark tests, the array operated successfully at 275 mK. The first cryogenic run with the windows and filters installed was conducted in April, 2018 with both the 250 and 350 \um FPAs installed. The optical loading on the detectors did not affect the operating temperature, allowing for preliminary optical testing. 

\begin{figure}[ht]
\begin{center}
\includegraphics[width=\linewidth]{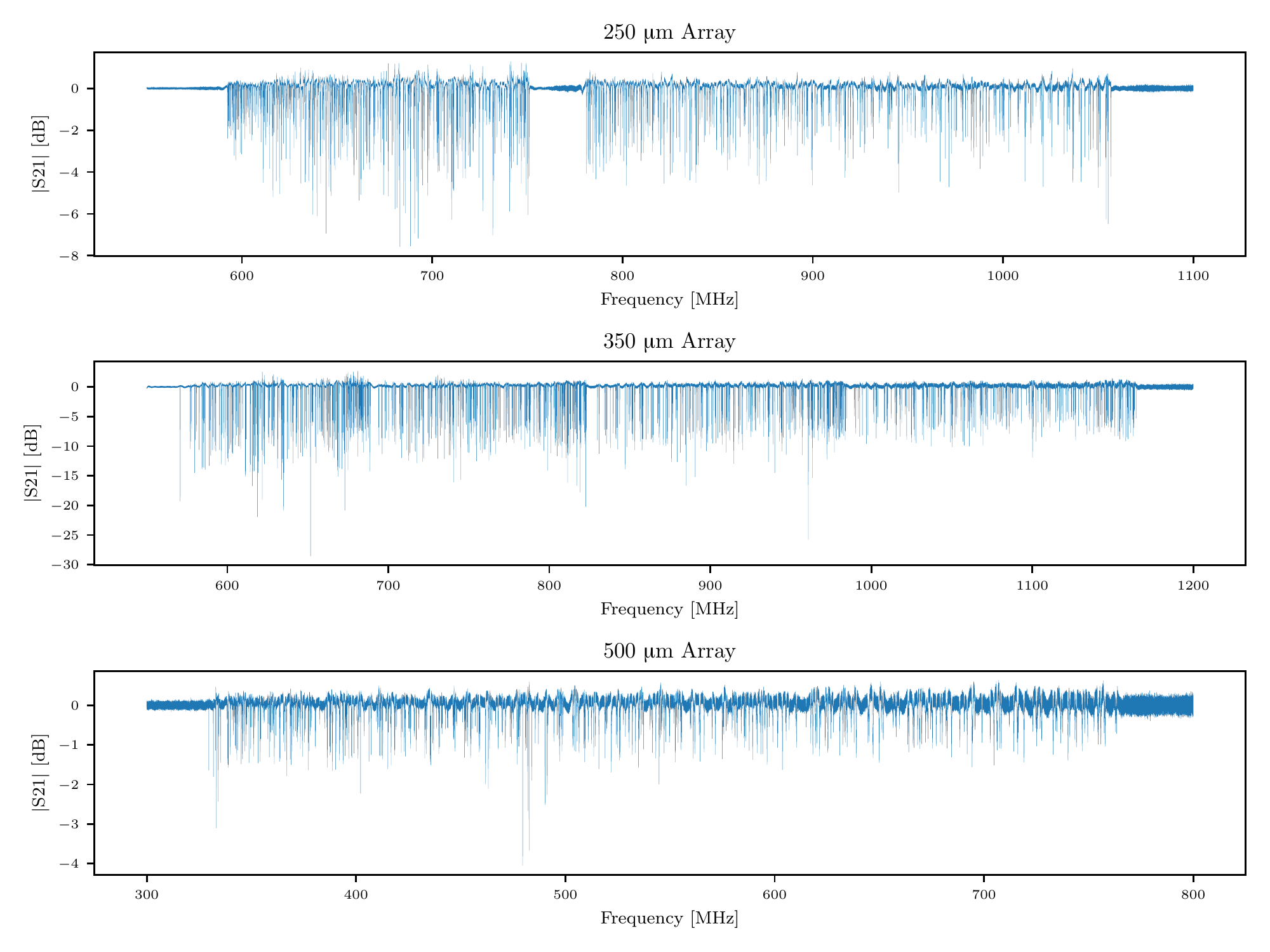}
\caption[VNA Sweeps for Each Array] {VNA sweeps for each of the arrays taken in the BLAST-TNG flight receiver. Data has been high-pass filtered to remove ripple and slope from cable attenuation. The sweep shown from the 250 \um array is for one of the three identical subarrays of MKIDs that make up the full FPA.}
\label{figure:sweeps}
\end{center}
\end{figure}

As of early May 2018, all three MKID FPAs are mounted in the receiver in flight configuration. Initial results indicate that all arrays are fully operational, with high detector yield and expected sensitivity levels. Vector network analyzer (VNA) sweeps for each array taken in the flight receiver are shown in Fig. \ref{figure:sweeps}.

\section{Flight Receiver Optical Testing}
\label{section:testing}  

\subsection{Polarization Response}
\begin{figure}[ht!]
\centering 
\includegraphics[width=.9\linewidth]{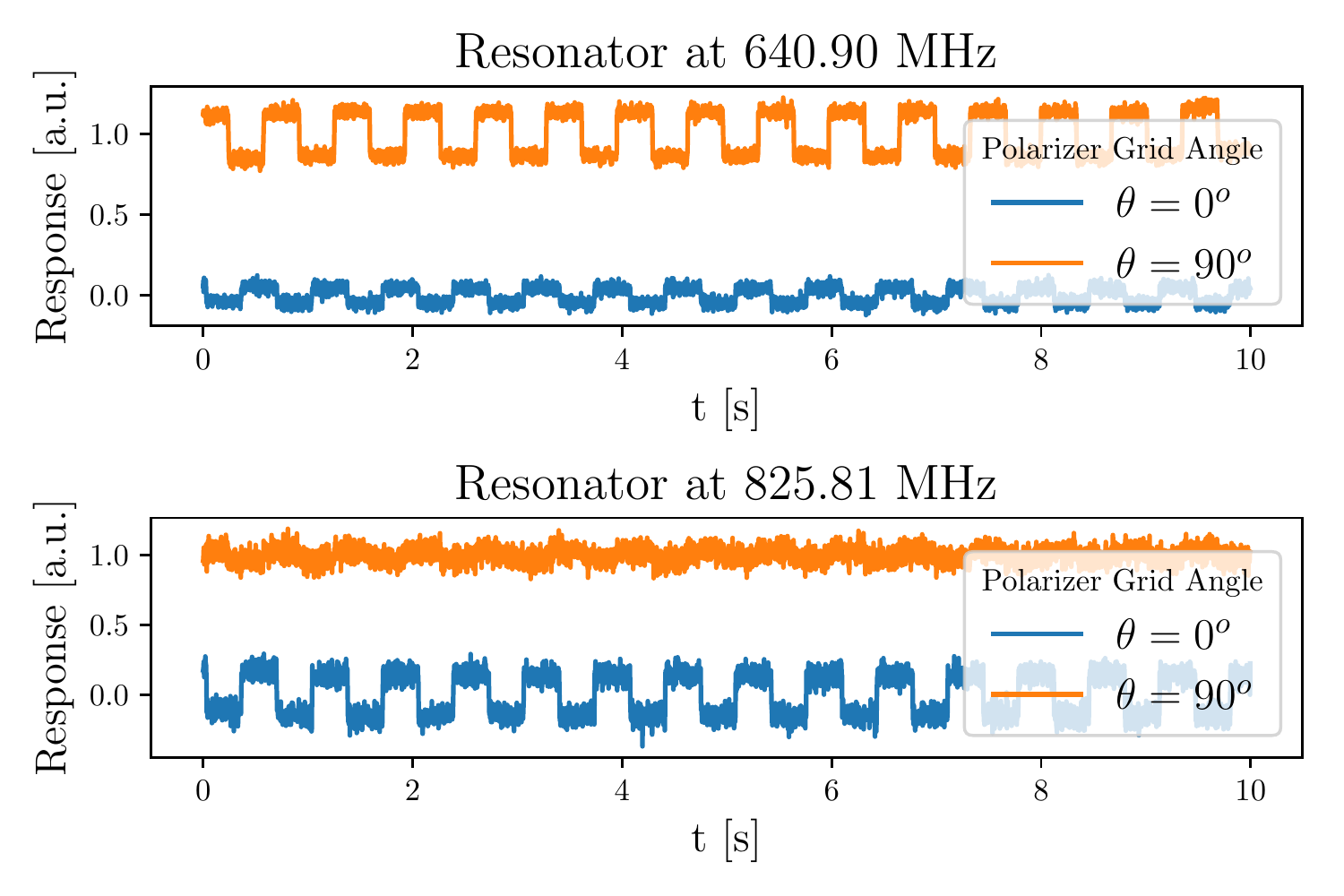}
\caption[Polarization Response Demonstration]{Preliminary polarization response demonstration of the 250 \um FPA, showing response to a 300 K/77 K chop viewed through a polarizer grid at different angles. Upper plot shows response of an X-Pol resonator and the lower plot shows a Y-Pol. Response is in arbitrary uncalibrated units read in using the ROACH electronics in flight configuration. Chop time streams are arbitrarily offset along the y-axis for better visibility. Polarizer grid absolute angles are arbitrary and are not referenced to the detector antenna axis.}
\label{figure:pol_eff}
\end{figure}

Understanding the response of the receiver to polarized light is critical to characterizing the instrument and analyzing the maps made during the flight. The detector cross-polar coupling has been measured at NIST to be less than 3\% \cite{brad_kids}. The polarization efficiency and cross-polar response of the receiver is measured by observing the response to a chopped thermal source which provides a near-square-wave chop between 300 K and 315 K. The source  fills the receiver beam and provides a uniform signal across most of the array. To ensure the detectors operate when viewing a 300 K blackbody we place a 4 K 4\% neutral density filter (NDF) at the entrance to the optics box. A metal-mesh polarizing grid provided by Cardiff University is placed in front of the cryostat and mounted at a 45\deg angle to the optical axis. The angled grid passes linearly-polarized light parallel to the grid orientation, and reflects perpendicularly-polarized light to an absorber outside the cryostat. Detector time streams are recorded using the flight ROACH-2 readout hardware during chops and the angle of the polarizer grid is stepped to measure the response as a function of polarization angle. Detector time streams measured during preliminary measurements are shown in Fig. \ref{figure:pol_eff} for representative X and Y-polarized resonators on the 250 \um array. Initial results, without the HWP installed, indicate a maximum cross-polar coupling of 4-6\% across all three arrays.

The degree of instrumentally-induced polarization signal in the receiver is measured by using the 4 K half wave plate (HWP) in the receiver cold optics. Instrumental polarization is characterized by repeating the same observations of the chopped thermal  source with the external grid at a fixed angle while stepping the HWP. Repeating these measurements at different grid angles allows polarization effects inherent to the receiver design to be identified and accounted for during data analysis. Initial polarization characterization was done with the HWP removed from the system, while improvements were made to the heat-strapping of the HWP rotator. The HWP has been reinstalled in the system, and will undergo full characterization during June, 2018.

\subsection{Noise}
The BLAST-TNG detector arrays have been demonstrated to be photon-noise-limited at the expected in-flight optical loading \cite{hubmayr_kid_photon_noise}, and a number of improvements have been made to the receiver RF system to maintain this low-noise performance. In order to minimize the conductive thermal loading on the liquid helium bath and the two VCS, we use thin (0.86 mm diameter) stainless steel coaxial cable \footnote[1]{COAX CO. LTD., Kanagawa, Japan} between 4 K and 300 K. These cables have sufficiently low thermal conductivity, but have relatively high signal attenuation, contributing to a round-trip signal attenuation of $\sim$ -30 dBm. 

Maintaining high signal-to-noise detector operation requires sufficient cold amplification, which is achieved with 4 K SiGe low-noise amplifiers designed by Arizona State University. Increasing the bias power of the amplifiers increases their gain, but also increases the load on the liquid helium bath and reduces the hold time of the cryostat. By cascading two SiGe amplifiers we achieve sufficient gain and demonstrated that the comparing the white noise level (away from the MKID resonant frequencies) of the cold amplifiers exceeds that of the warm readout electronics.

Initial tests of the cascaded amplifier scheme with the 250 \um array indicate that the detector noise level exceeds that of the full readout chain. By comparing the white noise level on and off resonance we have demonstrated that the system is detector-noise limited across most of the readout band. Though work is still ongoing to optimize the readout tone power for each detector, we expect the receiver and readout architecture to achieve photon-noise-limited performance in flight. 

\begin{figure}[ht!]
\centering
\includegraphics[width=.8\linewidth]{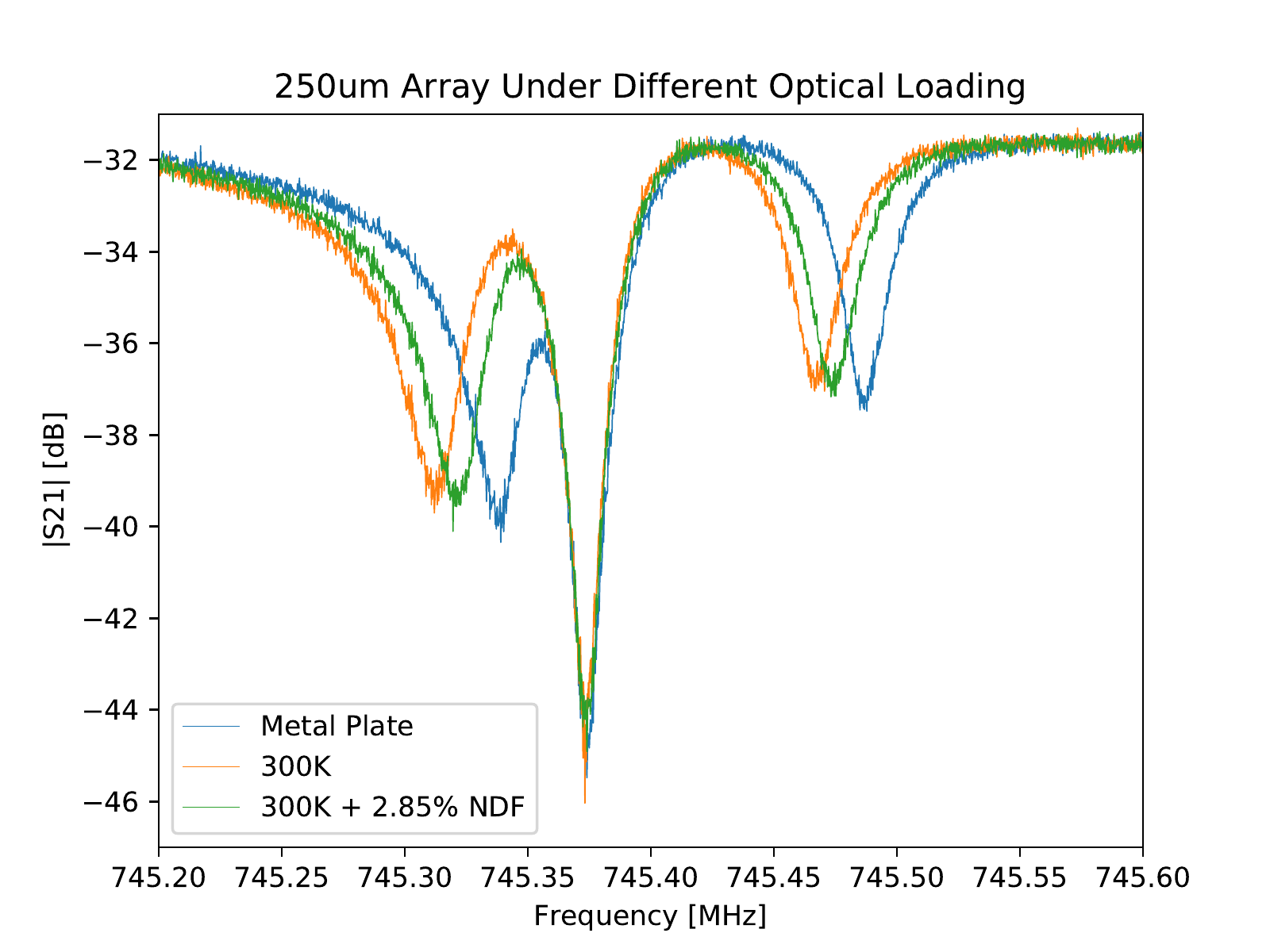}
\caption[250um Array Responsivity]{Response of three MKIDs on the 250\um array to different beam-filling optical loads, measured through  the BLAST-TNG cryogenic receiver with a vector network analyzer. The three curves show a different optical load: a metal plate covering the window of the receiver (blue), a 300 K blackbody (orange), and a 300 K blackbody viewed through a 2.85 \% neutral density filter (NDF). The resonator towards the center of the figure is a dark pixel (non-optically-coupled), while the resonators on either side of the figure are feedhorn-coupled to the cold optics. As the optical power is increased, the resonant frequencies and quality factors of the optically-coupled resonators decreases. The dark pixels show no change with the optical load. There is a clear difference in the quality factor and resonator depth between the dark and light pixels, due to the optical loading. We expect the optical loading in flight to be significantly reduced compared to these laboratory tests.}
\label{figure:responsivity}
\end{figure}

\subsection{Responsivity}

The sensitivity of the receiver is determined by measuring the detector response through the full cold optical system to known optical loads. For these measurements to be accurate, the optical load must fill the entire beam of the receiver. The same chopped optical source used in the polarization measurements was stepped in 5 K steps from 300 K to 330 K. At each temperature, the resonant frequencies of the MKIDs were determined using the ROACH-2 readout in vector network analyzer (VNA) mode, and the response to a 1 Hz chop between the room temperature and hot blackbody loads was recorded. Additional data was recorded viewing a 300 K blackbody through a room-temperature 2.85\% neutral density filter (NDF), and viewing a 77 K liquid nitrogen source. The response of approximately a dozen dark (not-feedhorn-coupled) MKIDs on each array can put limits on the cross-talk between pixels. During these laboratory tests, a 4.0\% NDF was placed at the entrance to the 4 K optics box to reduce the optical load and ensure the detectors operated when viewing the 300 K thermal load. The response of several resonators is shown in Fig.\ref{figure:responsivity}. Full characterization of the response of each array is still ongoing.

\subsection{Spectral Response}
The bandpass of each FPA will be measured with a Fourier transform spectrometer (FTS). An FTS, designed to operate at the BLAST-TNG wavebands, has been built at the University of Pennsylvania in collaboration with Cardiff University, and is currently being tested using the BLAST-TNG receiver. These measurements are ongoing, and we expect to finalize the bandpass measurements before summer 2018.

\section{Conclusion}
\label{section:conclusion}  

BLAST-TNG will be the most sensitive submillimeter polarimeter to date, and will build on the observing techniques developed for BLAST-Pol to make deeper maps of more science targets at higher resolution. The cryogenic receiver, MKID arrays, and ROACH-2-based readout are operating in flight configuration. Array level optical characterization data is still being analyzed, but initial results indicate the detector arrays in the flight receiver will maintain the photon-noise-limited performance demonstrated during single-pixel testing. The receiver will be integrated with the balloon gondola and telescope during pre-flight systems integration at the NASA Columbia Scientific Ballooning Facility in Palestine, TX, in preparation for a planned 28 day stratospheric balloon flight from McMurdo Station, Antarctica during the winter of 2018/2019.

\acknowledgments 
 
The BLAST-TNG collaboration acknowledges the support of NASA under award numbers NNX13AE50G and 80NSSC18K0481, and the NNX13CM03C. Detector development is supported in part by NASA through NNH13ZDA001N-APRA. J.D.S. acknowledges the support from the European Research Council (ERC) under the Horizon 2020 Framework Program via the Consolidator Grant CSF-648505. S.G. is supported through a NASA Earth and Space Science Fellowship (NESSF)  NNX16AO91H. The BLAST-TNG telescope is supported in part through the NASA SBIR/STTR office and developed at Alliance Spacesystems. The BLAST-TNG collaboration would like to acknowledge the Xilinx University Program for their generous donation of five Virtex-6 FPGAs for use in our ROACH-2 readout electronics. The collaboration also acknowledges the extensive machining, design, and fabrication efforts of Jeffrey Hancock and Harold Borders at the University of Pennsylvania and Matthew Underhill at Arizona State University, as well as Richard Gummer and the team at Precision Cryogenics Inc, without whose dedication the development and subsequent rebuild of the receiver would not have been possible. The BLAST-TNG team also recognizes the contribution of undergraduate and post-baccalaureate interns to the receiver development, especially Mark	Giovinazzi, Erin Healy, Gregory Kofman, Ariana Martino, Aaron Mathews, Timothy McSorely, Michael Plumb, and Nathan Schor, and Stephen Russell who built the calibration chopper used in the receiver testing.

\bibliography{report} 
\bibliographystyle{spiebib} 
\end{document}